\documentclass[12pt]{article}
\usepackage{epsfig}
\usepackage{slashed}
\usepackage{wrapfig}

\def\beq{\begin{equation}}
\def\eeq#1{\label{#1}\end{equation}}
\def\eeqn{\end{equation}}

\def\beqa{\begin{eqnarray}}
\def\eeqa#1{\label{#1}\end{eqnarray}}
\def\eeqan{\end{eqnarray}}

\let\bar=\overbar

\def\Dslash{\not{\hbox{\kern-4pt $D$}}}
\def\dslash{\not{\hbox{\kern-2pt $\del$}}}

\def\msb{{\bar{\ssstyle M \kern -1pt S}}}

\def\Title#1{\begin{center} {\Large {\bf #1} } \end{center}}

\begin{document}

\Title{Double Parton Scattering Contribution\\ to pp $\rightarrow$ $W^+W^+$}

\bigskip\bigskip


\begin{raggedright}

{\it Miroslav Myska\index{Myska, M.}\\
Department of Physics\\
FNSPE, Czech Technical University in Prague\\
Brehova 7, 115 19, Prague, Czech Republic}
\bigskip\bigskip
\end{raggedright}

\section{Introduction}

With larger energies of hadron collisions, the sea component of parton distribution
functions (PDF) grows rapidly and the probability for multiple parton interactions
(MPI) within the same collision becomes non-negligible, see e.g. \cite{PRD81_052012}. This study searches for the kinematic 
selection criteria for estimation of measurable MPI fraction in p-p collisions at
$\sqrt{s}$ = 14 $TeV$. The process under study is the production of pair of $W^+$ gauge bosons decaying 
leptonically into a pair of same-sign muons. 

\section{Signal Selection}

Four types of single parton scattering (SPS) background processes are studied 
in order to distinguis the kinemacics of double parton scattering (DPS) signal process
from the total same-sign di-muon data. These processes include $W^+W^+jj$, $W^+Z$, $ZZ$, and $t\bar{t}$
productions. The strongest selective power can be found in jet analysis and in detection of oppositely charged muons.
All events containing at least one jet with $p_T > 20$ GeV or events containing $\mu^-$ with $p_T > 5$ GeV are rejected.
Further signal selection can be reached via detailed muon kinematics analysis. Results are written in Table \ref{tab:sigma_summary}.
The detector acceptance and transverse momentum thresholds are motivated by the ATLAS detector performance \cite{ATLAS}. 
There are other tools allowing

\begin{table}[h!]
\begin{center}
\begin{tabular}{l | c | c c c c}
$\sigma~[fb]$ &DPS: $W^+W^+$ &SPS: $W^+W^+jj$ & $W^+Z$ & $ZZ$ & $t\bar{t}$ \\
\hline
$All~events$           & 1.96 & 4.59 & 68.21 & 36.41 & 8.8$\times10^3$ \\

$Full~Event~Selection$  & 0.86 & 0.04 & 1.85 & 0.27 & 0.89              \\

\footnotesize ~~~~survived    & \footnotesize 44\% & \footnotesize 1\%  & \footnotesize 3\%  & \footnotesize 1\%  & \footnotesize $10^{-2}$\%       \\
\hline
\end{tabular}
\caption{Summary of five studied processes characterized by production cross sections as well as by the fractions of appropriate events
surviving the final selection.}
\label{tab:sigma_summary}
\end{center}
\end{table}

\begin{wrapfigure}[12]{r}{4cm}
\begin{center}
\vspace{-0.5cm}
\includegraphics[width=4cm]{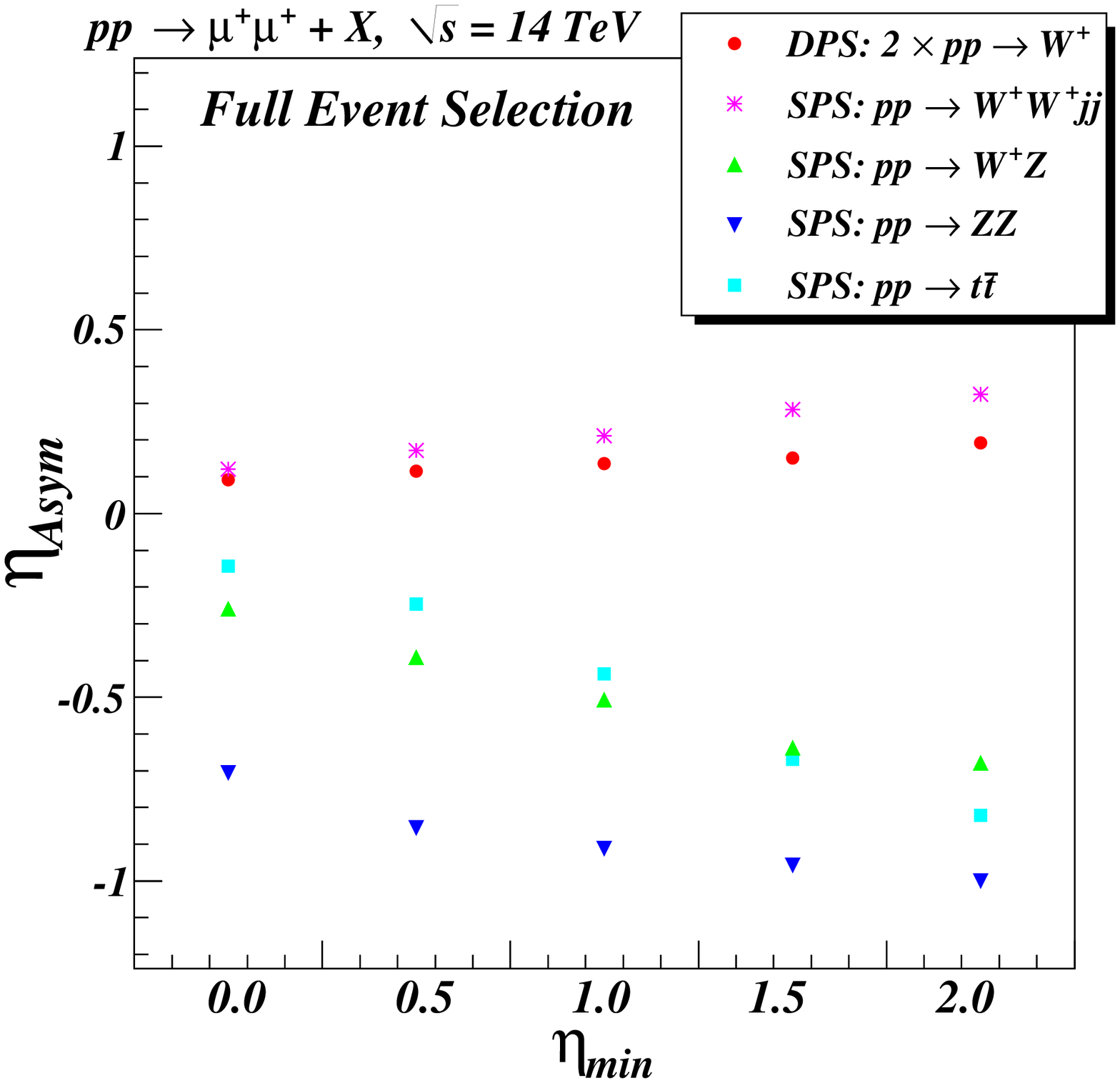}
\caption{Results for muon $\eta$ asymmetry.}
\label{fig:LepPair_Sel1_EtaAsym}
\end{center}
\end{wrapfigure}

\noindent signal separation but their usage has to be carefuly considered. The signal production cross section 
is already very small and the statistics should be kept as high as possible.

For instance, pseudo-rapidity distributions can be used to separate muons
into two detector hemispheres according to its sign. Both DPS and
SPS $W^+W^+$ processes slightly prefer to have the muons in the
opposite $\eta$ hemispheres ($\eta_{Asym}>0$), while the $W^+Z$, $ZZ$, and $t\bar{t}$
processes prefer muons occupying the same $\eta$
hemispheres ($\eta_{Asym}<0$). Figure \ref{fig:LepPair_Sel1_EtaAsym} shows the
$\eta_{Asym}$ values for the individual data sub-samples in five
bins. In order to explore the maximum selective power of this
asymmetry, we plot its values in dependence on the minimal allowed
pseudo-rapidity of muons, $\eta_{min}$, as suggested e.g. in \cite{EPJC69_53}. 
The bin for $\eta_{min} = 0$ corresponds to the entire considered detector acceptance. Higher bins
cut gradually off the central region of the detector in order to increase the difference among the
data sub-samples. We may conclude that the more forward muons,
the more significant differences in $\eta_{Asym}$ values for the
investigated processes can be observed.

\section{Summary}

The analysis of kinematical characteristics of single and double parton scattering 
was performed using the LO Monte Carlo generators Herwig++ \cite{Herwig} and MadGraph \cite{MadGraph}.
The final result will strongly depend on the detection and identification efficiencies of the specific detector.
However, the parton level results presented in this paper indicates the measurability of the studied process
with the estimated cross section of 0.84 fb. Signal-to-background ratio of 0.28 leads to the high luminosity
measurements. $\mathcal{O}(100fb^{-1})$ of data is required to be analysed in order to reach signal significance of $5\sigma$ above the background.

\end{document}